# Longitudinal structural connectomic and rich-club analysis in adolescent mTBI reveals persistent, distributed brain alterations acutely through to one year post-injury

Ai Wern Chung[1*], Rebekah Mannix[2], Henry A. Feldman[3], P. Ellen Grant[1], Kiho Im[1*]

[1]Fetal Neonatal Neuroimaging and Developmental Science Center, Division of Newborn Medicine, Boston Children's Hospital, Harvard Medical School, MA, USA

[2]Division of Emergency Medicine, Brain Injury Center, Boston Children's Hospital, Harvard Medical School, MA, USA

[3]Institutional Centers for Clinical and Translational Research and Division of Newborn Medicine, Boston Children's Hospital, Harvard Medical School, MA, USA

*Corresponding authors: `aiwern.chung@childrens.harvard.edu`; `kiho.im@childrens.harvard.edu`

**Abstract**

The diffused nature of mild traumatic brain injury (mTBI) impacts brain white-matter pathways with potentially long-term consequences, even after initial symptoms have resolved. To understand post-mTBI recovery in adolescents, longitudinal studies are needed to determine the interplay between highly individualised recovery trajectories and ongoing development. To capture the distributed nature of mTBI and recovery, we employ connectomes to probe the brain's structural organisation. We present a diffusion MRI study on adolescent mTBI subjects scanned one day, two weeks and one year after injury with controls. Longitudinal global network changes over time suggests an altered and more 'diffuse' network topology post-injury (specifically lower transitivity and global efficiency). Stratifying the connectome by its back-bone, known as the 'rich-club', these network changes were driven by the 'peripheral' local subnetwork by way of increased network density, fractional anisotropy and decreased diffusivities. This increased structural integrity of the local subnetwork may be to compensate for an injured network, or it may be robust to mTBI and is exhibiting a normal developmental trend. The rich-club also revealed lower diffusivities over time with controls, potentially indicative of longer-term structural ramifications. Our results show evolving, diffuse alterations in adolescent mTBI connectomes beginning acutely and continuing to one year.

**Keywords**: adolescents, mild traumatic brain injury, concussion, network theory, rich club

**Introduction**

Mild traumatic brain injury (mTBI) or concussion poses an immense public health burden, particularly in adolescents. Adolescents accounted for greater than 750/100,000 of Emergency Department visits between 2002-2006 in the US[1]. While deemed "mild", concussed adolescents remain symptomatic for more than two weeks in 50% of cases and up to one month in 30% of children after injury[2]. Despite this large public health burden, clinicians have few objective tools to guide diagnosis and management of this common adolescent injury. As such, there has been increasing interest in the use of neuroimaging for objective markers of injury to improve patient monitoring and to better understand the mechanisms of concussion on a neurobiological level[3–5]. However, most efforts have focused on adults, and such findings may not be translatable to adolescents due to differences in neurobiological properties such as maturity of tissue (myelination), greater vulnerability to impact (weaker neck musculature and support), increased water content and propensity towards cerebral edema[6,7]. In addition, neural developmental processes are ongoing in the adolescent brain, alongside the capacity for recovery through compensatory or developmental mechanisms[7,8]. While symptoms of neurologic dysfunction post-concussion are transient, diffused mechanical injury may have longer-lasting implications on brain structure given that the time-course for physiological recovery exceeds current clinical clearance[9]. For these reasons, we present an adolescent study that is longitudinal in design, vital for understanding the longer-term impact of concussion in this age group.

Advanced diffusion-weighted MRI and its related techniques have revealed localised, microstructural white-matter injury in mTBI[3] but the relationship of focal findings to overall brain function as an organised system remains poorly characterised. Structural neuroimaging analyses are shifting towards a global interrogation of the brain as an inter-connected system using *network theory*[10–13]. In network theory, the cerebral cortex is parcellated into regions that are defined as *nodes*. Nodes are connected by *edges* defined by tractography reconstructed white-matter fibres, where edges are weighted by features of connectivity strength such as the number of fibres between nodes[14]. This graph of the brain incorporates localised, structural alterations and includes their contribution towards global, whole brain measures that capture the network's organisation or 'topology'[15]. Given the distributed nature of concussive injury and the ensuing biochemical cascade (focal versus diffused injuries, primary versus secondary responses and injuries over time)[6,16] analysing diffusion-based, structural data with a *gestalt* approach afforded by network theory may be appropriate to identify *disorganisation* or aberrations in the mTBI brain. There are few diffusion network theory studies on children and adolescents with strictly mild TBI[4,17]. To date, one study in children found significant diffusion network differences between mTBI in the acute stage (MRI within 72 hours post-injury) and controls, specifically a reduction in global integration and greater regional segregation in the network[18]. For children in the chronic stage of TBI (from complicated mild through to severe injuries), significant network differences similar to acute mTBI[18] have been found up to nine years post-injury[19–21].

In addition to global network theory measures, *rich-club* based analysis is a complementary investigation of subnetworks in relation to a collection of highly inter-connected *hub* brain regions. Hubs are critical for integrating distributed network domains and possess characteristics (such as having a high number of connections and high metabolic demand) potentially reflective of their importance for efficient neuronal signalling and communication in the brain[22–24]. Densely connected hubs form a "rich-club" subnetwork that has been consistently identified in studies across ages in both normal and patient populations[25–31]. The rich-club is a high-cost network in terms of wiring and metabolic usage, making them equally vulnerable in a number of disorders[13,23]. It has exhibited impaired structural connectivity in adolescent chronic moderate to severe TBI[19] and increased functional connectivity in adult subacute and chronic mild to severe TBI[32,33]. To date, the characteristics of rich-club connectivity in adolescent mild TBI have not been studied, certainly not longitudinally, and is one of the primary contributions of this work.

Brain structure in adolescence continues to change into adulthood[34], leading to brain networks that differ from adults[35]. Network theoretical studies in children and adolescent concussion have been cross-sectional[18–21]. Our pilot study investigates longitudinal changes of global structural network organisation in adolescents with mTBI assessed at the acute (< 3 days post-injury), subacute (two weeks post-injury) and chronic stages (one year post-injury), in relation to controls. We sought to further understand any observed changes in network topology alongside its rich-club organisation to identify subnetworks which may explain an altered system in adolescent concussion over time.

**Material and Methods**

*Subjects*
This study was approved by the Institutional Review Board at Boston Children's Hospital. Informed consent was obtained from all participants and/or their legal guardians and research was performed in accordance with relevant guidelines/regulations. The patient group presented to the Emergency Department within 24 hours of injury (n=9). Each mTBI subject was scanned longitudinally within 72 hours (acute), two weeks (subacute) and one year (chronic) post-injury. mTBI was defined as a blunt, sports-related injury to the head resulting in either (1) alteration in mental status (including loss of consciousness, disorientation, or amnesia) or (2) any of the following symptoms that started within four hours of injury and were not present before the injury: headache, nausea, vomiting, dizziness/balance problems, fatigue, drowsiness, blurred vision, memory difficulty or difficulty concentrating. Patients were excluded from the study if they presented to the Emergency Department with Glasgow Coma Scale<14, focal symptoms or other indications for head imaging or intracranial hemorrhage seen when imaging was obtained, orthopaedic fracture, co-existing intra-abdominal or intra-thoracic trauma, or spinal-cord injury, or an underlying neurologic disorder or psychiatric illness requiring medications. A group of healthy controls were also recruited and MRI scanned once (n=9), matched for age and sex with the patient group at the acute time-point. Controls were recruited without current neurological complaints or recent head trauma at least a year prior to scanning.

*MRI Acquisition*

T1- and diffusion-weighted imaging (DWI) data were acquired on a 3T Siemens Tim Trio system, maximal gradient strength 40mT/m (Erlangen, Germany). T1-weighted motion mitigated multi-echo MPRAGE[36] parameters were: TR=2520ms; TE=1.74, 3.54, 5.34 and 7.14ms; inversion time=1350ms; FOV=240mm$^2$; voxel size=1mm$^3$. DWI simultaneous multi-slice, echo-planar imaging[37] parameters were: 63 non-collinear gradient direction volumes acquired at $b$=3000s/mm$^2$; 4 $b$=0 s/mm$^2$; TR=5800ms; TE=119ms; FOV=240mm$^2$, voxel size=2mm$^3$. Additional non-diffusion weighted volumes acquired in the anterior-posterior and posterior-anterior direction were obtained (one for each phase-encoding direction) for susceptibility artefact correction in the pre-processing stage later.

*Image Processing*

T1-Weighted Data - were pre-processed with the Freesurfer 'recon-all' pipeline (https://surfer.nmr.mgh.harvard.edu) and output were visually assessed. The T1-weighted cortical surface was parcellated into cortical regions as defined by the Desikan-Killiany atlas[38]. A transform between native skull-stripped b0 to T1-weighted space was computed via a rigid and affine-registration (NiftyReg, http://cmictig.cs.ucl.ac.uk/wiki/index.php/NiftyReg[39]). The inverse of this transform, t', was used to register the Freesurfer cortical parcellation to $b$0 space. t' was also applied on Freesurfer's whole brain white-matter mask to map it to b0 space for tractography.

DWI Data - were visually assessed for artefacts and up to three corrupt gradient volumes were removed for each dataset. Remaining DWI data were corrected for susceptibility artefact, subject motion and eddy current distortion with topup and eddy in FSL[40,41]. Gradient *b*-vectors were rotated following eddy. Linear least squares diffusion model fitting and 2nd order Runga-Kutta tractography was achieved using Diffusion Toolkit and visualised with TrackVis[42]. Tractography was seeded from the centre of each white-matter voxel in the brain, with a tracking step-size of 0.1mm and angular threshold of 45°. Tracts with length between 20mm and 200mm were retained. The cortical regions reached by each tract's endpoint was recorded. The number of streamlines connecting any two regions formed the corresponding entry in the connectome. Network nodes included 68 cortical regions, and edge weights were the number of streamlines connecting pair-wise nodes. Connectivity matrices were normalised by the total number of tracts in the same matrix[30]. The following imaging and network features were calculated for all mTBI subjects' time-point data and for controls.

*Whole Brain Diffusion Measures Analyses*

We determine the potential and extent of whole brain diffusion measures and volume change during this developmental phase of our subjects to assess whether these changes may impact the networks constructed. Whole brain white-, cortical and deep grey-matter tissue volumes were calculated using T1-weighted images from their respective Freesurfer masks. Whole brain tissue volume was calculated by combining all three masks. Each tissue mask was mapped to native diffusion space (with the t' transform, as described in 'Image Processing) and the FA and mean diffusivity (MD) values within each mask were averaged.

*Network Analyses*

Global Network Theory Analyses

Analyses presented in this manuscript were performed on weighted networks. Network theoretical measures calculated (https://sites.google.com/site/bctnet) were: network density, global transitivity, efficiency, node betweenness-centrality, characteristic path length, modularity, clustering coefficient and the small-world coefficient[15]. We generated 1000 random realisations of the observed network while preserving network size, degree distribution and density[43]. A network measure is then normalized by dividing it by the mean corresponding measure computed across all random realisations to assess if the determined differences are random. We computed these null-network normalised measures for transitivity, characteristic path length, global efficiency and clustering coefficient. This is commonly done to improve the comparability of determined network measures between groups and subjects[14,15,18,19]. Figure 1 shows the stages from network construction onwards. To investigate the effect of the diffusion model on network theory measures we repeated this analysis on connectomes computed from QBall reconstructed data (see Supplementary Analysis S1 for full Methods).

Rich-Club Connectivity Analyses

Rich-club (RC) nodes were defined as ten *a priori* regions on the Desikan-Killiany Freesurfer atlas: the superior frontal and parietal gyri, precuneus, posterior cingulate and insular bilaterally[26,44–47]. These regions have been consistently established as key brain hubs across age, pathologies and species[13,22,28,48,49]. To confirm that these *a priori* regions were reasonable for our population, traditional RC analysis was also performed on controls and the patient group at each time-point, on both weighted and binary networks (see Supplementary Analysis S2). Normalised RC coefficients[50,51] were computed on group-level connectomes, revealing significant values greater than 1 for a range of degrees (Figure S2.1). The majority of RC nodes extracted based on these analyses intersected with all the above *a priori* regions (Tables S2.1-S2.3). For the experiment presented for the remainder of this manuscript, three subnetworks were defined by grouping edges according to their association with the *a priori* RC nodes: 'rich-club' (connecting RC nodes only), 'feeder' (connecting an RC and a non-RC node) and 'local' (connecting non-RC nodes only) subnetworks. Figure 1 illustrates how a model network is divided into subnetworks in relation to rich-club nodes, and plots the connectomes of the three subnetworks for a control subject. The network density (number of connected edges, NoE), average number of reconstructed streamlines per edge (NoS), average FA, MD, axial and radial diffusivities (AD and RD) were calculated for each subnetwork for statistical analyses.

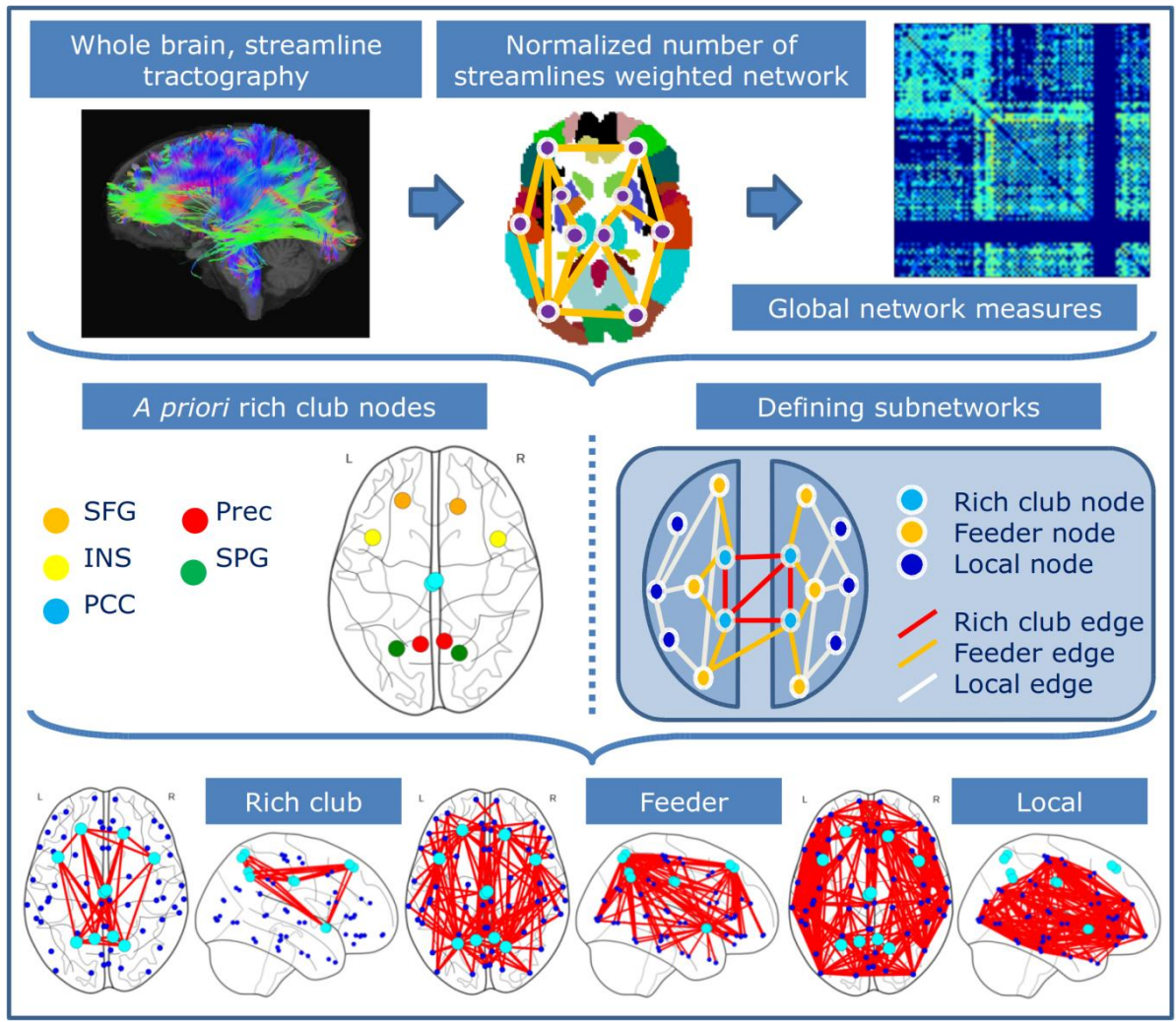

**Figure 1 - Overview of network construction and rich-club subnetwork definitions. Top panel:** depicts general post-processing steps from tractography. **Middle panel:** Using *a priori* definitions of ten bilateral rich-club nodes (SFG = superior frontal gyrus, INS = insular, PCC = posterior cingulate, Prec = precuneus, SPG = superior parietal gyrus), edges and remaining nodes in the network are subsequently grouped into rich-club, feeder and local subnetworks according to their relationship to the rich-club nodes. **Bottom panel:** Given the *a priori* rich-club nodes (in cyan), the axial and sagittal projections of each subnetwork is plotted for an example Control subject.

*Statistical Analyses*

Fisher's exact test was used to compare sex differences between control and mTBI groups, and Mann-Whitney U-tests to compare their ages. The following statistical analyses were performed on brain volume and diffusion measures, network theoretical measures and diffusivity measures from RC analyses, using SPSS (IBM, SPSS Statistics, 2012). ANCOVA covarying for age at time of MRI was performed for cross-sectional comparisons between controls and mTBI subjects at the acute and chronic time-point. For longitudinal mTBI group analyses, a repeated-measures ANCOVA model was performed with the following equation fitting the outcome measure $y_{it}$ for subject *i* at time-point *t*=[1,2,3] : $y_{it} = a + (b \times age_i) + c_t + e_{it}$, with main effect time-point representing acute, subacute and chronic stages, and $e_{it}$ is the error term with mean 0 and variance $\delta^2_t$ (specific to time). Time-point was treated as a 3-level factor to enable nonlinear trends with time. Analyses were covaried by subject's age at baseline MRI (acute time-point for mTBI or time of MRI for control subjects). An unstructured covariance matrix was employed to allow for arbitrary correlations for a given subject's dependent variable over time. This model enabled the use of all data points by accounting for missing time-points. In the interest of understanding longitudinal effect of concussion, we limit our report of results to significance found with time-effect. Where only main factor 'time-point' is significant, the observed changes are significant irrespective of a subject's age. If both 'time-point' and 'age at baseline MRI' factors are significant, then the age at time of concussion has an effect on the MRI measure but not on the observed change with time. As such, results with only significant age-effect are reported in Supplementary Materials. We took *p* <= 0.05 as the criterion for statistical significance. Due to the exploratory nature of our pilot study, *p*-values are uncorrected for multiple comparisons correction.

**Results**

*Demographics*

Table 1 contains group demographics of subjects analysed and the MRI timings for each mTBI time-point. Following visual inspection of the MRI data, two mTBI subjects in the subacute stage and one control were removed from analyses due to excessive motion during acquisition. Another two mTBI subjects were only scanned at a single time-point. This resulted in five mTBI subjects with full longitudinal data available at all three time-points, two subjects with data at both acute and chronic time-points, and two subjects with data from a single time-point (one subject acquired at the acute stage, and another subject at the subacute stage). The sex and age between control and mTBI groups were non-significant (all *p*>0.05). Every effort was made to recruit controls age-matched to mTBI patients in the acute time-point, with Mann-Whitney U-tests revealing a closer match of controls to patients at the acute time-point (control vs. acute (*p*-value = 0.879); control vs. chronic (*p* = 0.152)). On average, patients were scanned 1, 17 and 418 days post-injury. See Supplementary Table S3 for all ages and data analysed at each time-point.

**Table 1 - Demographics and time of MRI**

|  | Controls | mTBI Patients |
|---|---|---|
| **Number of subjects [Acute/Subacute/Chronic]** | 8 | 8/6/7 |
| **Gender F/M [Acute][Subacute][Chronic]** | 1/7 | [1/7],[1/5],[1/6]# |
| **Median (IQR) [range] age at time of scan** | | |
| **Controls** | 13.46(1.62) [11.97 - 15.84] | |
| **mTBI Patients:** | | |
| Acute | | 13.34(1.54) [11.51-20.29]* |
| Subacute | | 13.32 (3.50) [12.42-20.33]* |
| Chronic | | 14.36 (2.36) [12.65-21.41]* |
| **Mean (stdev) [range] time of scan post-injury** | | |
| Acute (days) | | 1.1 (0.7) [0-2] |
| Subacute (days) | | 16.8 (2.2) [5-20] |
| Chronic (years) | | 1.15 (0.06) [1.10-1.26] |

#Fisher's test on sex differences between groups all $p > 0.05$.

*All $p > 0.1$ for age comparisons between Controls versus each mTBI time-point (Mann-Whitney U test).

IQR = Inter-quartile range.

*Whole Brain Diffusion Measures Analyses*

Full statistical results and figures of volume and diffusion measures for each whole brain segmentation can be found in Supplementary Figure S3 and Table S4. Cross-sectionally, there was a significant increase in deep grey-matter FA for mTBI chronically compared to controls (Figure 2) (MD was conversely significantly lower, Figure S3). Although not significant, mTBI subjects in the acute stage exhibited, on average, greater FA than controls in white- and deep grey-matter (Figure 2a and 2c). White-matter volume was significantly larger at chronic time-point versus controls. Longitudinally, mTBI subjects had a significant change in deep grey-matter FA with time ($F(2, 6.85) = 8.806$, $p = 0.013$) with greater FA at the chronic stage compared to previous time-points (Figure 2c). Consistent with FA increases, a significant decrease in MD was also found in deep grey-matter ($p = 0.001$, Figure S3, with age-effect also significant). All remaining whole brain regions did not show significant FA, MD and tissue volume changes over time (all $p > 0.05$).

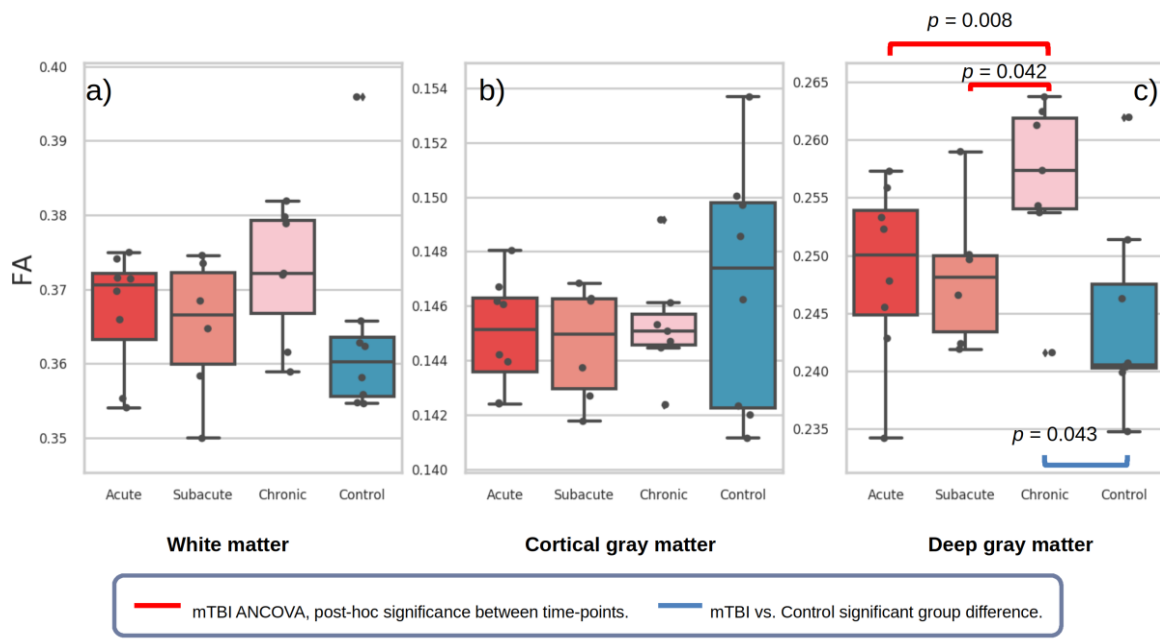

**Figure 2** - Box plots of FA in whole brain tissue masks for each mTBI time-point and Controls, with individual data points plotted as black circles. Significant statistical result for cross-sectional *t*-tests between mTBI stage and controls are denoted in blue. Longitudinal general linear model results for mTBI are represented in red brackets, where significant post-hoc differences between time-points are shown when the corresponding ANCOVA analysis is significant. Age at MRI was not a significant confounder. Outliers are denoted at 1.5 times the inter-quartile range by black diamonds. Outliers are similarly defined in all further boxplot figures.

*Global Network Theory Analyses*

Network theoretical measures for each group, full statistical results and box plots are in Supplementary Figure S4 and Table S5. Cross-sectionally, there was a significant decrease in global efficiency ($p = 0.05$) in chronic mTBI subjects compared to controls (Figure 3b, in blue). No significance was reached for other global networks between controls and mTBI.

More significant results were found longitudinally within the mTBI cohort, specifically in transitivity ($F(2, 2.865) = 21.884$, $p = 0.018$), global efficiency ($F(2, 6.263) = 14.764$, $p = 0.004$) and network degree ($F(2, 3.975) = 11.232$, $p = 0.023$) (Figure 3, in red). Transitivity and global efficiency decreased significantly over time: transitivity was significantly greater in the acute stage compared to both remaining time-points (Figure 3a); and global efficiency was significantly lower in the chronic time-point compared to earlier time-points (Figure 3b). Degree increased significantly over time, being significantly lower in the acute stage compared to later time-points (Figure 3c). Normalised measures of transitivity and global efficiency were similarly significant and trending (Supplementary Figure S4, and Table S5). Although not significant, it is of interest to note qualitatively the 'normalisation' of several network measures over time longitudinally when comparing acute with chronic time-points, ending with values within the ranges of controls (small-world coefficient, shortest path measures, see Supplementary Figure S4).

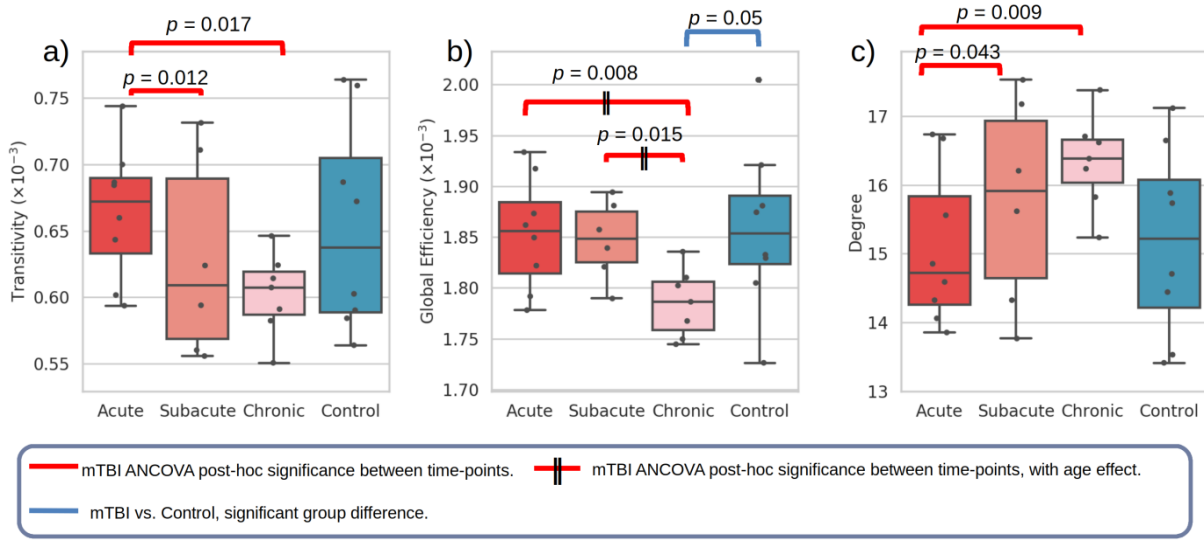

**Figure 3 - Box plots of global network theoretical measures transitivity, global efficiency and degree at each mTBI stage and for controls. Longitudinal general linear model for mTBI are represented in red, where significant post-hoc differences between time-points are shown when the corresponding ANCOVA analysis is significant.**

Repeating the analysis on multi-fibre model, QBall, revealed largely similar network measure trends over time and between Controls and mTBI as with DTI (namely betweenness centrality, transitivity, global efficiency, modularity and degree. See Analysis S1 and Figure S1.1). Cross-sectional comparisons between mTBI and Controls were largely non-significant and transitivity and global efficiency measures exhibited significant changes over time for mTBI subjects in both models (Table S1.1). It is possible that in future work with larger sample sizes, the significance of these trends will be more consistent between models.

*Rich-Club Connectivity Analyses*

Figure 4 shows box plots of each network and diffusion measure grouped by cohort and subnetwork. See Supplementary Materials Table S6 for full statistical results. Longitudinally, mean NoS in the RC subnetwork is significantly altered over time (Figure 4a, in red, $F(2, 4.701) = 8.283$, $p = 0.029$). However, most significant longitudinal changes were observed in the local subnetwork, namely in network density (NoE) ($F(2, 5.662) = 5.315$, $p = 0.05$), FA ($F(2, 5.264) = 9.658$, $p = 0.017$), MD ($F(2, 7.130) = 4.990$, $p = 0.044$) and RD ($F(2, 6.147) = 7.424$, $p = 0.023$)  (MD and RD were also significant for age-effect) with most differences occurring between the acute stage and remaining two time-points (Figure 4 and Table S6). Although not significant, mean NoS was greater in the mTBI cohort at all stages compared to controls in the RC networks (Figure 4a). This increase in NoS would most likely be due to the greater whole brain FA (compared to controls, see Figure 2) leading to more tracts being reconstructed. FA is also observed within each subnetwork (Table S6) to significantly increase over time in the local subnetwork, and remained relatively consistent in RC and feeder networks. These diffusion changes in the network may explain the rise in connected edges in more 'peripheral' subnetworks (NoE, feeder $F(2, 3.509) = 8.250$, $p = 0.047$, local $F(2, 5.662) = 5.315$, $p = 0.05$). An interesting observation are the non-linear 'elbow' trends in many of the measures and subnetworks from acute to chronic time-points (Figure 4 and Supplementary Material Table S6), most being significant in the local subnetworks.

Cross-sectionally, MD was significantly lower in acute versus controls in the RC subnetwork ($p = 0.045$, Figure 4c) which was primarily driven by decreasing RD (for both acute and chronic time-points versus controls at $p = 0.039$ and $0.027$, respectively, Figure 4e), Although not significant, for all three subnetworks, mean FA was greater in mTBI acute subjects compared to controls, with all remaining diffusivity measures (MD, AD and RD) behaving similarly in the opposite direction as expected (lower diffusivity in mTBI compared to controls).

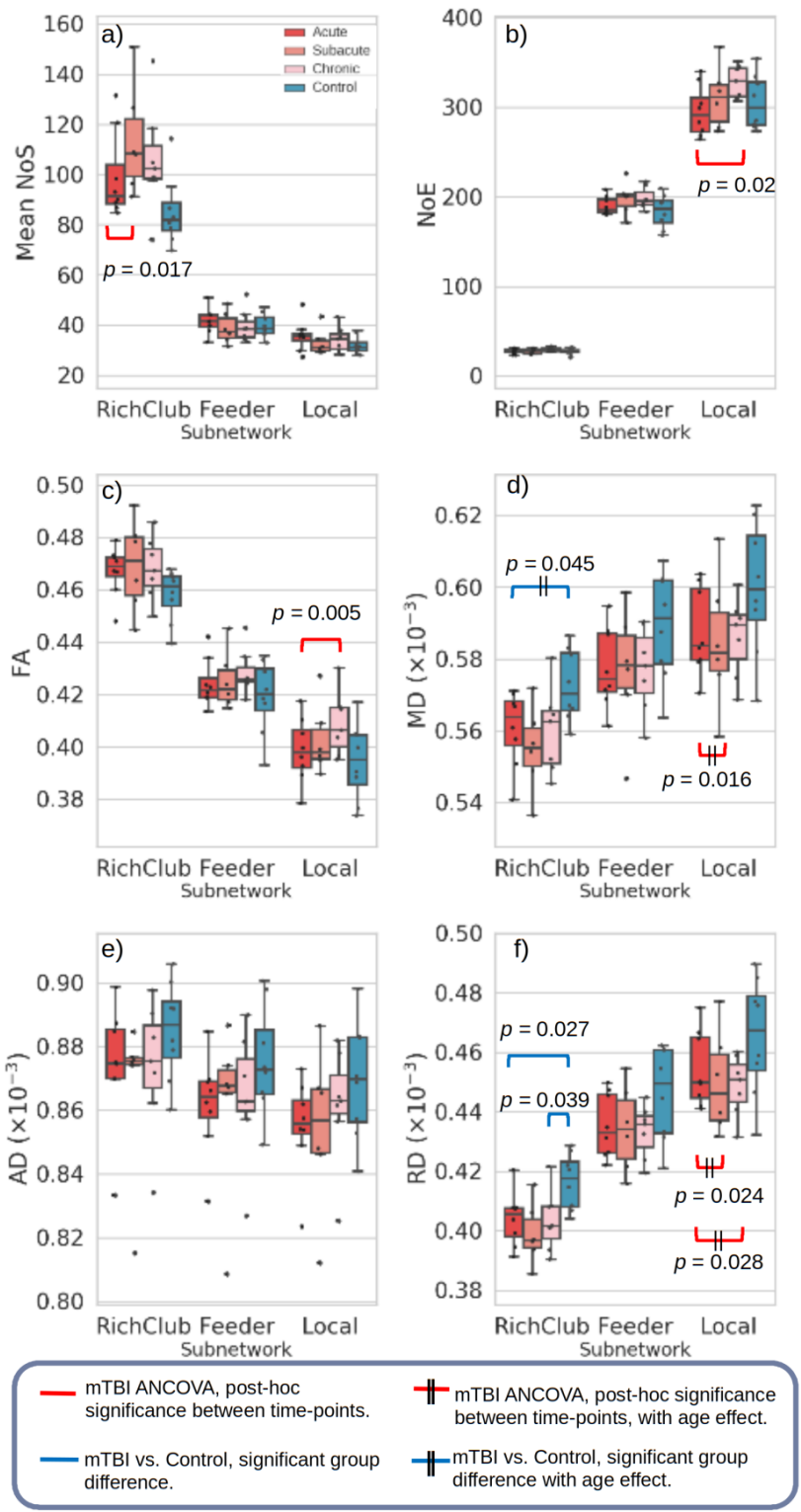

**Figure 4 - Boxplots of mean diffusion measures for each subject group, in each subnetwork (rich-club, feeder and local subnetworks). Significant statistical result for cross-sectional *t*-tests between controls and mTBI stages are denoted in blue. Longitudinal general linear model for mTBI are represented in red, where post-hoc differences between time-points are shown when the corresponding ANCOVA analysis is significant. NoS = Number of streamlines, NoE = Number of edges.**

**Discussion**

To our knowledge, this is the first network theoretical and rich-club investigation on the longitudinal trajectory of brain structural organisation in concussed adolescents. Patients were scanned acutely (< 3 days), subacutely (2 weeks) and chronically (1 year) after injury. Our work also addressed the lack of mTBI diffusion MRI studies in adolescents, with the aim to utilise a global model of analysis offered by connectomics to understand the distributed nature of concussion injury. Compared with controls, the rich-club was the only subnetwork with significantly lower diffusivity measures in mTBI subjects at acute and chronic time-points. Longitudinally, we found alterations in global network topology indicative of lower efficiency over time that may be due to diffusion changes in the local subnetwork of the brain.

*Global Network Theory Analyses*

We found few significant difference in network metrics between mTBI and control groups. Unlike Yuan *et al*, we did not observe significant change in the acute stage compared to controls, although comparing medians suggests trends similar to theirs (namely increased small-world coefficient and transitivity)[18]. With the addition of more subjects in our pilot, these trends could become significant. With respect to differences in controls and chronic mTBI, we found significant decrease in global efficiency without a significant impact on other network metrics. Our chronic time-point of one year post-injury is earlier than two other studies in adolescent mTBI, which ranged from one to nine years post-injury[20,21]. These studies found an increase in small-world coefficient and also largely no significant network metric changes[20]. It may be that beyond a year after injury, global network features are similar to controls in adolescent mTBI, unlike in moderate to severe cases where network alterations have been documented from one to six years post-injury[19,21]. Interestingly, in our longitudinally analysis, patients revealed overall decreased transitivity and global efficiency, suggestive of a reduction in both network segregation and integration over time. Without other longitudinal network analyses in adolescent mTBI, we can only discern that while significant differences with controls are weak, we are able to detect within-subject network alterations up to a year post-injury. Whether these changes remain beyond a year remains to be studied.

*Rich-Club Connectivity Analyses*

We found significant structural changes in all three subnetworks in contrast with controls and also longitudinally within our mTBI cohort. Only the rich-club subnetwork revealed significant differences with controls - with primarily lower diffusivity in mTBI soon after injury and one year later. Longitudinally significant changes also occurred in the subnetworks, namely the rich-club subnetwork in terms of mean NoS and the local subnetwork exhibiting the most alterations over time (in number of edges, FA, MD and RD). Our analyses suggest that significant FA increase in the local subnetwork drove the increase in network density observed in this same subnetwork.

A rich-club analysis in adolescent moderate to severe TBI found differential patterns of change in connectivity strength between subnetworks in relation to the rich-club, when compared to controls[19]. Two years post-injury, Verhelst *et al.* identified increased connectivity in the local subnetwork which they suggest may be a compensatory effect to their observed decrease in rich-club connectivity. However, the subjects in Verhelst *et al.* presented with distributed diffuse axonal injury on MRI at time of injury with Glasgow Coma Scale 3 to 7, were comatose for >24 hours to 7 weeks and had encephalomalacia in the majority of cases. Interestingly, we found alterations in rich-club based subnetworks for our cohort which suffered from a milder form of brain injury. Theoretically the rich-club is the backbone of a network and is therefore by nature of its high inter-connectivity robust to network failures (or injury) as it provides alternative pathways to maintain network stability[24]. This characteristic makes this subnetwork rigid and constrained with lower 'evolvability'[52], or biologically with respect to the brain, may possess lower plasticity, or too high a cost for repair. As the only subnetwork with enough sensitivity to display significant difference with controls, the rich-club's response to mTBI may also be a gradual process, affecting recovery or normal development beyond that of a year (compared to the more flexible peripheral subnetworks). This lack of sufficient time to detect recovery or developmental changes may explain why we did not find significant longitudinal trends in the rich-club for concussion subjects. Taken all together, our results suggest subnetworks are altered with changes found as early as three days after concussion, with potential aberrations through to one year particularly in the local subnetwork. Furthermore, the 'diffuse' organisation observed from our global network analyses may be explained by a shift in "importance" from the rich-club to the local subnetwork as measured by their diffusion properties.

*Subnetwork Analyses and the Biomechanics of Concussion*

But what does it mean for a local subnetwork to be altered? In contrast to the high metabolic and wiring cost of the rich-club, the more 'peripheral' local subnetwork requires less up-keep[23] and perhaps its strengthening is indicative of a mechanism to compensate for an affected rich-club. What is yet to be established is not only a biomechanical explanation for the differential changes in strength and connectivity between subnetworks in the brain, particularly over time after injury, but simply the role of feeder and local subnetworks for communication in the brain and their corresponding neuro-anatomy. Identifying a node as a rich-club member is dependent on how connected it is, which is in turn associated with its corresponding edges possessing high FA to drive tractography. Thus the rich-club subnetwork typically consists of high FA tracts that are highly myelinated for effective action potential and information propagation, whereas feeder and local subnetworks are predominantly comprised of lower FA pathways[23]. The high rotational acceleration forces from concussion on the brain have been extensively modelled to include both FA and the directional information from the diffusion tensor to model tissue 'stiffness' and capture the strain from impact[53–56]. Such models not only show deformation to be widely distributed in white-matter throughout the brain, but suggest that regions with high fiber directionality (therefore high FA) have greater maximal principle stresses than regions with lower FA[54,56]. It is reasonable to hypothesise from this that the local subnetwork may be less likely (compared to the rich-club) to be directly affected following an impact. This may also explain why the rich-club was the most sensitive subnetwork in exhibiting significant diffusivity changes when compared to controls. Moreover, perhaps the local subnetwork was less affected and is presenting significant changes that are expected in normal development (increased FA, decreased MD) over the course of a year in our adolescent population. Whether our observed changes to the local subnetwork are simply due to normal development or are compensatory in nature remains to be investigated with a longer follow-up period and neuropsychological assessment.

Significant differences between controls and mTBI in diffusion measures of FA and MD are commonly reported in cross-sectional studies employing localised region-of-interest or tract-based image analyses. Typically, greater FA at the acute stages and lower FA in the chronic stage is found when compared to controls[3,4,17]. These group differences in specific regions do not appear to manifest as a global deficit in terms of global network measures when compared with controls. Even so, global network measures exhibit sensitivity on an individual level by way of greater longitudinal significance, demonstrating the importance of longitudinal analyses and assessment in mTBI. Interestingly, analysing the network by rich-club associated subnetworks revealed similar diffusivity trends between controls and mTBI in the acute and chronic stages as in other region-based DTI studies[3,4]. In addition, we observed non-linear trends across mTBI time-points in these subnetworks, possibly capturing a period of normalisation in our network and diffusion measures, potentially returning to pre-injury values or transiently passing through normal[57–59].

*Technical Considerations*

Our mTBI cohort exhibited the general trend of increased white-matter volume and FA, and decreased MD that typifies adolescent development[60], however, none of these changes reached significance (see *Whole Brain Diffusion Measures Analyses*). Only deep grey-matter FA increase and MD decrease reached significance with grey-matter tissue volumes remaining unchanged over time. The longitudinal stability of white-matter measures suggests that normal developmental changes over one year in our cohort are unlikely to impact tractography results and the networks computed (particularly as we excluded deep grey-matter regions from our network). In particular, our observations with age are unlikely to be driven by significantly larger white-matter volumes or greater FA producing more streamlines. To further account for normal developmental changes in our analyses, we normalised our networks according to total number of streamlines for each subject and statistically factored for age to account for the age range in our mTBI group.

In addition to the complexity of comparing concussion studies with different types of injury and stages of recovery when measurements are made, technical differences arising from network theoretical analysis must also be considered. There is no consensus on network construction in the neuroscience community to date with differences in diffusion modelling, tract reconstruction, choice of nodes, edge weights, network normalisation procedure and rich-club node definition all contributing toward variations in TBI network theoretical findings in literature[61]. One variation we account for is by using *a priori* regions to define the rich club, regions that have been repeatedly reproduced in literature and similarly in our cohorts (Supplementary Analysis S2). The rich club regions identified largely overlapped between weighted and binary networks and between cohorts, with greatest variation between weighted and binary networks, demonstrating the ability of tract information to restrict the rich club to a small number of similar regions that are biologically plausible[23,24,62]. For an in-depth discussion on the effect of weighted versus binary networks on the rich-club, see Supplementary Analysis 2.3. Using these *a priori* RC nodes ensures consistency when comparing across our groups and time-points, and also with other studies. Fixing the rich club to a set of *a priori* regions is also particularly important in our subsequent analyses, such as for feeder and local subnetwork analysis. A point of consideration is our use of the diffusion tensor model as it is constrained to model diffusivity in a single fibre population, therefore unable to model multiple fibre populations with more complex configurations. By repeating our 'Global Network Theory Analyses' on connectomes derived from a multi-fibre diffusion model (Qball) we found largely similar network measure trends over time as the DTI model (Figure S1.1). There were fewer significant results in the QBall analysis, although measures that were significant were similar to DTI (transitivity, global efficiency). Future work with larger sample sizes will determine the significance of the observed trends and their consistency for and between models. The tensor has the advantage of extracting the most probable connections, thus limiting the inter-subject variability and reconstruction of false-positive tracts found in more complex models[63,64]. We also acquired high diffusion sensitisation MRI data at $b$=3000 s/mm$^2$ with other studies acquiring at $b$=750 to 1200 s/mm$^2$ [18–21]. High $b$-value imaging captures slow diffusing water molecules such as those associated with myelinated white-matter with greater sensitivity, enabling the detection of more subtle alterations in diffusion properties than lower $b$-values[65–67].

*Limitations*

The lack of significant differences compared to the control group may be due to the number of subjects in our pilot study being too small to overcome inter-subject variance in our groups. It should be noted that the network theoretical studies in children and adolescents discussed above are also of modest size, between 16 and 23 subjects[18–21]. Future work with larger sample sizes will confirm the findings reported here, elucidating whether subnetwork changes are from normal development or compensatory and whether our observations at one year after injury have normalised or are transient, alongside a longitudinal control cohort.

**Conclusion**

We investigated the longitudinal impact of mTBI in adolescent structural network organisation from the acute, to subacute and finally chronic stages after concussion. The evolution of mTBI networks revealed global changes in network specialisation and integration over time. Rich-club analysis suggested these global alterations may be driven by changes to the structural integrity of the local, 'peripheral', subnetwork. This may be a compensatory response to injury, or reflective of normal developmental maturation in the local subnetwork. Furthermore, the rich-club was the only subnetwork to have significantly lower diffusivities when compared to controls. Overall, our study points towards mTBI having a diffused and distributed effect on brain structural network organisation up to a year after injury. When these neurophysiological alterations resolve and the longer-term neurological implications remain to be determined. However our early investigation suggests continual patient observation may be necessary during this period of ongoing development.

**Acknowledgements**

The authors wish to thank the families that participated and our colleagues at Boston Children's Hospital for their help. This study was partially funded by the Harvard Catalyst, and the American Heart Association and Children's Heart Foundation 19POST34880005.

**Author Contributions**

R.M. and P.E.G. designed and supervised data collection. A.W.C. and K.I. were responsible for imaging analysis design. A.W.C. conducted the experiments and analysed the results. H.A.F., R.M. and K.I. provided statistical analysis guidance. A.W.C. and P.E.G. wrote the paper, all authors reviewed the manuscript.